\documentclass[conference]{IEEEtran}
\IEEEoverridecommandlockouts
\usepackage{url}
\usepackage{cite}
\usepackage{amsmath,amssymb,amsfonts}
\usepackage{algorithmic}
\usepackage{graphicx}
\usepackage{textcomp}
\usepackage{xcolor}
\usepackage{booktabs}
\usepackage{caption}
\def\BibTeX{{\rm B\kern-.05em{\sc i\kern-.025em b}\kern-.08em
    T\kern-.1667em\lower.7ex\hbox{E}\kern-.125emX}}
\begin{document}

\title{Efficient Image Compression Using Advanced State Space Models}


\author{\IEEEauthorblockN{Bouzid Arezki}
\IEEEauthorblockA{\textit{L2TI Laboratory} \\
\textit{University Sorbonne Paris Nord}\\
Villetaneuse, France \\
bouzid.arezki@edu.univ-paris13.fr}
\and
\IEEEauthorblockN{Anissa Mokraoui}
\IEEEauthorblockA{\textit{L2TI Laboratory} \\
\textit{University Sorbonne Paris Nord}\\
Villetaneuse, France \\
anissa.mokraoui@univ-paris13.fr}
\and
\IEEEauthorblockN{Fangchen Feng}
\IEEEauthorblockA{\textit{L2TI Laboratory} \\
\textit{University Sorbonne Paris Nord}\\
Villetaneuse, France \\
fangchen.feng@univ-paris13.fr}

}

\maketitle

\begin{abstract}
Transformers have led to learning-based image compression methods that outperform traditional approaches. However, these methods often suffer from high complexity, limiting their practical application. To address this, various strategies such as knowledge distillation and lightweight architectures have been explored, aiming to enhance efficiency without significantly sacrificing performance. This paper proposes a State Space Model-based Image Compression (SSMIC) architecture. This novel architecture balances performance and computational efficiency, making it suitable for real-world applications. Experimental evaluations confirm the effectiveness of our model in achieving a superior BD-rate while significantly reducing computational complexity and latency compared to competitive learning-based image compression methods.

\end{abstract}


\section{Introduction}
Visual compression represents a fundamental challenge in multimedia processing. Over the past few decades, classical standards have predominantly been employed including BPG, JPEG, JPEG2000, H.265/HEVC, and H.266/VVC. With the introduction of deep neural network architectures such as Convolutional Neural Networks (CNNs)~\cite{balle2018variational,Learning_Convo,CondiProModel,NEURIPS2018_53edebc5,lee2018contextadaptive,9190935} and Transformers~\cite{Entroformer,swinT,stf}, learning-based compression methods have emerged, demonstrating continuously improving performance and garnering interest over traditional approaches. These architectures typically include a two-level hierarchical variational autoencoder with a hyper-prior as the entropy model. They consist of two sets of encoders and decoders: one for the generative model and another for the hyper-prior model. However, deep learning-based compression approaches often suffer from high complexity, which limits their practicality in real-world applications. Although it is common practice to propose scaled-down models by choosing a relatively smaller model size, these models, while less complex and faster, often significantly sacrifice the compression performance~\cite{swinT}. Other approaches have been deployed to accelerate these models while maintaining their performance. Knowledge distillation is one effective method for accelerating neural networks across various computer vision tasks~\cite{beyer2022knowledge}. In this context, EVC~\cite{guo2022evc} used a mask decay method based on teacher-student training. Another promising direction involves the development of lightweight  architectures~\cite{wu2023pslt,mehta2021mobilevit}. Recent advancements in lightweight attention mechanisms have resulted in architectures with improved inference times for image compression~\cite{he2024towards}. Other noteworthy contributions include~\cite{yang2023computationally}, where the authors focus on reducing decoding complexity through the proposal of shallow decoding transforms. Additionally, the work in~\cite{he2021checkerboard, ali2024towards} focuses on improving the efficiency of the entropy model.

\begin{figure}[h]
  \includegraphics[width=0.48\textwidth]{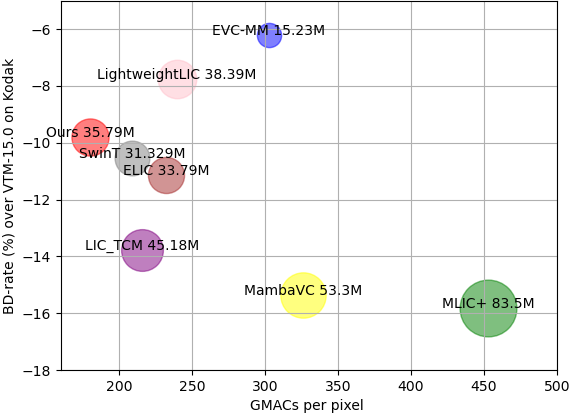}
  \caption{BD-rate performance over VTM-15.0~\cite{VTM} vs computational complexity (GMACs) on  Kodak~\cite{kodak}. The radius of the circles represents the number of parameters of the model. The further the method is positioned towards the bottom left of the graph, the better its performance. }
  \label{fig:bubble}
\end{figure}

Recently, the State Space model and its variant, the Mamba model, have garnered significant attention in the field of computer vision. State Space Models (SSMs) were initially introduced in deep neural networks for sequence modeling~\cite{ssm}, unifying the strengths of previous sequence models, including Continuous-Time Models (CTMs), Recurrent Neural Networks (RNNs), and CNNs. Despite their potential, SSMs have not seen widespread use due to their high computational and memory requirements, which stem the latent state representation connecting the input and output. Mamba~\cite{s4} addresses these limitations by integrating a selection mechanism into the structured variants of SSMs, enhancing their context-based reasoning ability. In this paper, we propose a State Space Model-based Image Compression (SSMIC) architecture focusing on rate-distortion performance, computational complexity, and latency. The contributions of this paper are as follows: 
\begin{itemize}
\item SSMIC balances compression efficiency, computational complexity, and latency for practical multimedia processing applications. 
\item SSMIC integrates SSMs from the Mamba model into the image compression pipeline, enhancing contextual reasoning while managing computational and memory requirements.
\item SSMIC is a lightweight architecture designed for efficient real-time image compression on resource-constrained devices.
\item Extensive experiments across benchmark datasets show that SSMIC achieves competitive compression performance with reduced computational complexity and latency compared to existing methods. Fig.~\ref{fig:bubble}, depicting the trade-off between BD-rate and computational complexity on the Kodak dataset, validates the standing of SSMIC among competitive state-of-the-art methods.

\end{itemize}

\section{Related works}
\subsection{Deep Learning-based Image Compression}
Among the early works, the authors of~\cite{balle2018variational,NEURIPS2018_53edebc5} introduced a CNN-based two-level hierarchical architecture. The authors of~\cite{Entroformer} incorporated an attention mechanism into the image compression framework by introducing self-attention in the hyper-prior model. A more sophisticated approach, the Swin block~\cite{liu2021swin}, has been proposed for use in both the generative and hyper-prior models~\cite{swinT}. Additionally, the authors of~\cite{TCM} proposed a Transformer-CNN Mixture (TCM) block, which leverages the strengths of both transformers and CNNs by splitting the feature dimension into separate branches for each. Although TCM improves compression performance compared to transformer-only architectures, it does not reduce encoding or decoding time.

\subsection{Lightweight Attention}
Ever since the attention mechanism for visual tasks was proposed and achieved impressive results, efforts to reduce its complexity have been underway. The Swin Transformer, for instance, introduced window-based self-attention to limit computational cost~\cite{swinT}. Mobile-Former~\cite{chen2022mobile} leverages the advantages of MobileNet~\cite{mobilenet} for local processing and the strengths of transformers for global interaction. Similarly, EdgeViT~\cite{pan2022edgevits} and EdgeNeXt~\cite{maaz2022edgenext} combine local interaction (convolution) with global self-attention. MobileViT~\cite{mehta2021mobilevit}, another lightweight, general-purpose vision transformer for mobile devices, uses a multi-scale sampler for efficient training. Additionally, LVT~\cite{yang2022lite} develops enhanced self-attention mechanisms to improve computational efficiency. A ladder self-attention and progressive shift mechanism was proposed in~\cite{wu2023pslt} for lightweight transformers. In \cite{zhang2022efficient}, an efficient long-range attention network was designed for image super-resolution tasks, which inspired \cite{he2024towards} for image compression. Other works  focused on image compression as well, such as \cite{SwinNPE}, which proposed a Swin-like block without positional encoding, thereby reducing the number of trainable parameters. 

\subsection{Mamba for Visual Compression}
The most closely related work to our approach is the recently proposed MambaVC~\cite{qin2024mambavc}, which incorporates the Mamba block into the image compression architecture to achieve promising results. Despite this similarity, our design differs significantly from theirs. While MambaVC focuses on improving compression performance, our approach is distinct in that it emphasizes designing a computationally efficient compression model. This difference in motivation results in unique design choices throughout the architecture, which will be detailed in the following section.

\section{Novel State Space Model-based Image Compression (SSMIC)}

\begin{figure*}[h]
\centering
  \includegraphics[width=0.935\textwidth]{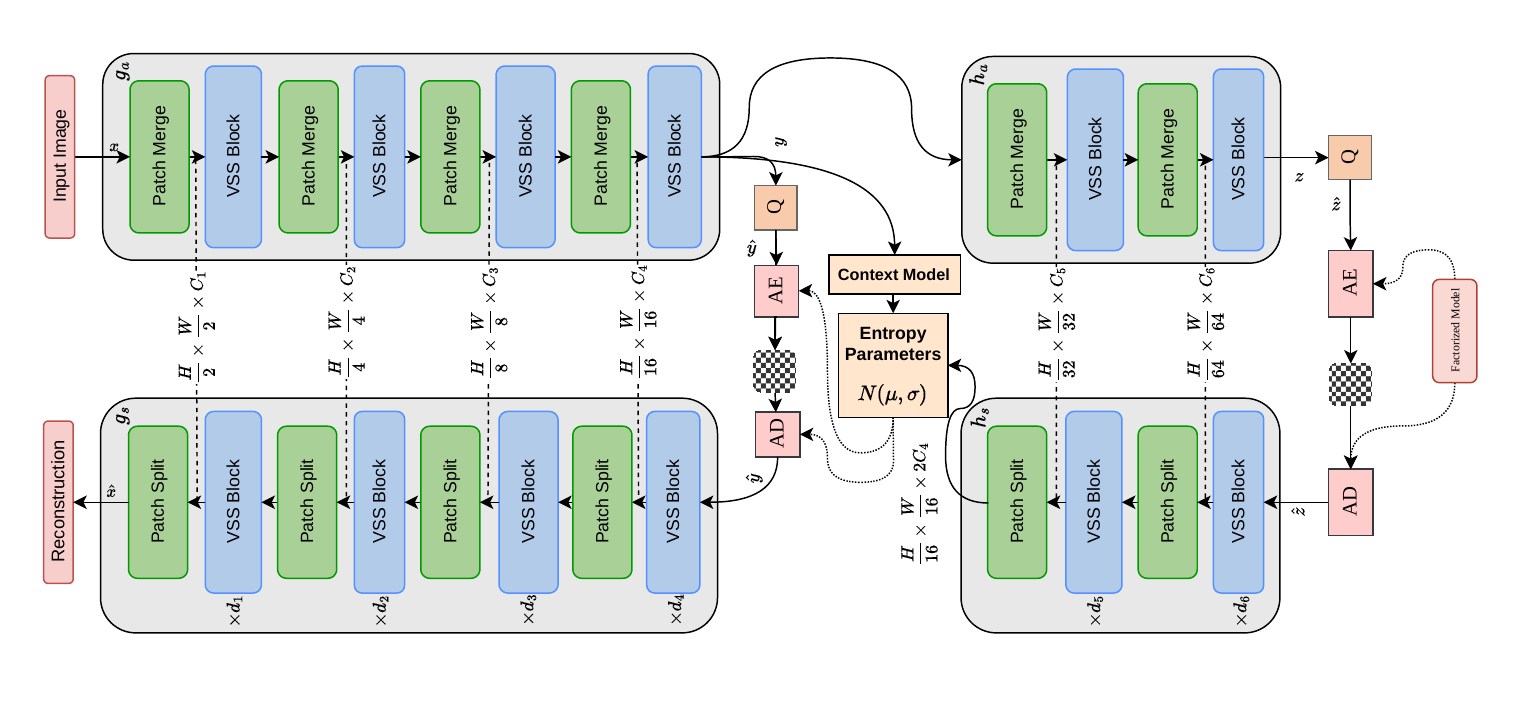}
  \caption{Network architecture of our proposed SSMIC model.}
  \label{fig:network}
\end{figure*}

\begin{figure*}[h]
\centering
  \includegraphics[width=0.82\textwidth]{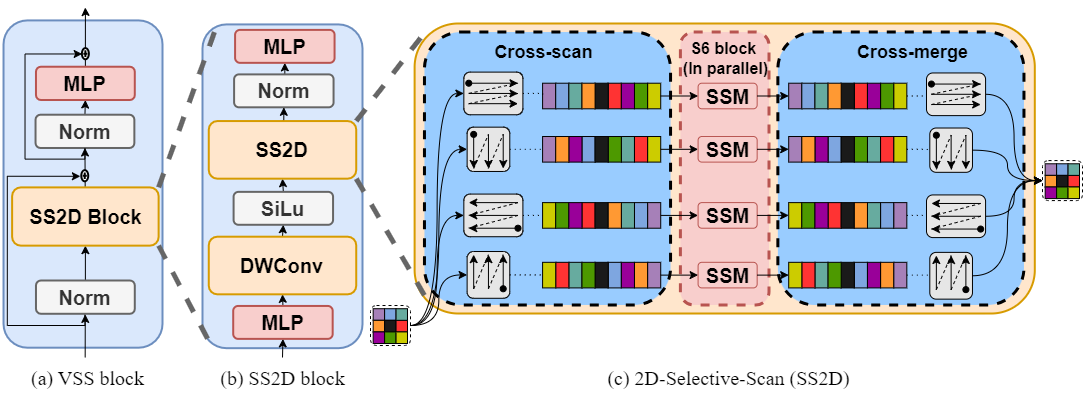}
  \caption{A VSS block \cite{vmamba} consists of an SS2D block, which performs selective scans in four parallel patterns.}
  \label{fig:vss}
\end{figure*}

\subsection{Preliminaries}
Before presenting our SSMIC architecture, we introduce some useful information hereafter.

The SSM transformation in S4~\cite{s4} originates from the classical state space model, which maps a one dimensional input signal $x(t) \in \mathbb{R}$ to a one dimensional output signal $y(t) \in \mathbb{R}$ through a latent state $h(t) \in \mathbb{R}^N$ of dimension $N$:
\begin{equation} \label{eq:ssm}
\begin{split}
&h'(t) = \mathbf{A}h(t)+ \mathbf{B}x(t), \\
&y(t) = \mathbf{C}h(t),
\end{split}
\end{equation}
where $\mathbf{A} \in \mathbb{R}^{N \times N}$, $\mathbf{B} \in \mathbb{R}^{N \times 1}$,
$\mathbf{C} \in \mathbb{R}^{1 \times N}$ are parameters of neural networks in deep learning.
To deal with the discrete input sequence $x= [x_0,x_1, \ldots, x_{L-1}] \in \mathbb{R}^L$, the parameters in equation~\eqref{eq:ssm} are discretized using a step size $\Delta$, which represents the resolution of the continuous input $x(t)$~\cite{s4}. In particular, the continuous parameters $\mathbf{A}$ and $\mathbf{B}$ are converted into discretization parameters $\overline{\mathbf{A}}$ and $\overline{\mathbf{B}}$ by the zero-order hold (ZOH) technique, defined as:
\begin{equation} \label{eq:ssm_discrete}
\begin{split}
&\overline{\mathbf{A}} = \exp(\Delta\mathbf{A}), \\
&\overline{\mathbf{B}} = (\Delta \mathbf{A})^{-1} (\exp(\Delta \mathbf{A})-\mathbf{I}).\Delta \mathbf{B}.
\end{split}
\end{equation}
After discretizing $\mathbf{A}$ and $\mathbf{B}$ to $\overline{\mathbf{A}}$ and $\overline{\mathbf{B}}$,  equation~\eqref{eq:ssm} can be reformulated as: 
\begin{equation} \label{eq:ssm_after_discretezation}
\begin{split}
&h_t = \overline{\mathbf{A}}h_{t-1} + \overline{\mathbf{B}}x_t, \\
&y_t = \mathbf{C}h_t.
\end{split}
\end{equation}
SSMs can be efficiently computed using RNNs. This recursive process can  be reformulated and computed as a convolution:
\begin{equation}
\begin{split}
&\overline{\mathbf{K}} = (\mathbf{C}\overline{\mathbf{B}}, \mathbf{C}\overline{\mathbf{A}\mathbf{B}},\ldots, \mathbf{C}\overline{\mathbf{A}}^{L-1}\overline{\mathbf{B}}), \\
&y=x*\overline{\mathbf{K}},
\end{split}
\end{equation}
where $L$ denotes the length of the input sequence $x$;  and $\overline{\mathbf{K}} \in \mathbb{R}^L$ is the SSM convolution kernel.

Mamba~\cite{mamba} further incorporates data-dependence to capture contextual information in equation~\eqref{eq:ssm} by proposing a novel parameterization method for SSMs that integrates an input-dependent selection mechanism, referred to as S6. Although the recurrent nature of SSMs restricts full parallelization, Mamba employs structural reparameterization tricks and a hardware-efficient parallel scanning algorithm to boost overall efficiency. As a result, many works have adapted Mamba from Natural Language Processing (NLP) to the vision domain~\cite{vmamba,liu2024vision}.

\subsection{Proposed SSMIC architecture}
The proposed SSMIC architecture follows a general two-level architecture. Specifically, the input image $x$ is first encoded by the generative encoder $y=g_a(x)$, and the hyper-latent $z=h_a(y)$ is obtained through the encoder of the hyper-prior network. The quantized version of the hyper-latent $\hat{z}$ is modeled and entropy-coded using a learned factorized model before being passed through $h_s(\hat{z})$. The output of $h_s$, along with the output of the context model, enters the entropy parameters network~\cite{NEURIPS2018_53edebc5}, which generates the mean $\mu$ and scale $\sigma$ parameters for a conditional Gaussian entropy model $P(y|\hat{z})= \mathcal{N}(\mu,\,\sigma^2)$ to model $y$. The quantized latent $\hat{y}=Q(y)$ is finally entropy-coded (Arithmetic encoding/decoding AE/AD) and sent to $\hat{x}=g_a(\hat{y})$ for reconstructing the image $\hat{x}$. A detailed illustration is shown in Fig.~\ref{fig:network}. 

\begin{table*}[t]
    \centering   
    \begin{tabular}{lcccc}
        \toprule
        Method & Kodak \cite{kodak} & CLIC2020 \cite{clic2020} & JPEG-AI \cite{jpegai}  & Average\\
        \midrule
        BPG444 \cite{bpg} & 29.86\% & 32.77\% & 43.77\% & 35.46\% \\
        SwinT* \cite{swinT} & -10.52\% & -6.47\% & -2.78\% &  -6.03\%\\
        MambaVC* \cite{qin2024mambavc}& -15.37\% & -16.69\% & -12.47\% & -14.69\% \\   
        SwinNPE \cite{SwinNPE}&  -5.85\% &  -17.50\% &  -23.56\% & -15.63\% \\
        LightweightLIC \cite{he2024towards}& -7.76\% & -23.60\% &  -29.86\% & -20.40\%\\
        ELIC \cite{he2022elic} & -11.14\% & -27.45\%& -31.31\% &-23.30\%\\
        MLIC+* \cite{mlic}&  \textbf{-15.86\%} &  - &  -15.89\% & - \\
        LIC\_TCM \cite{TCM} & -13.76\% & \textbf{-30.65\%}& \textbf{-33.37\%} &\textbf{-25.92\%}\\
        SSMIC (Ours) & -9.81\% & -29.91\% & -25.55\% & -21.75\% \\
        \bottomrule
    \end{tabular}
    \caption{BD-rate performance using  VTM-15.0 \cite{VTM} as reference. We use $*$ to indicate the approaches for which the numbers are sourced from their respective papers. The evaluation of MLIC+ on CLIC2020 dataset is not provided in~\cite{mlic}.}
    \label{tab:compression_comparison}
\end{table*}
We propose using the context model from~\cite{NEURIPS2018_53edebc5}, which draws inspiration by traditional coding techniques that predict the probability of unknown codes based on previously decoded latents. Unlike the architecture in~\cite{qin2024mambavc}, our choice is driven by the goal of developing a computationally efficient architecture. While more sophisticated channel-wise autoregressive models~\cite {9190935} can enhance performance, they are significantly more time-consuming~\cite{he2021checkerboard}. We use the classical strategy of adding uniform noise to simulate the quantization operation (Q) which makes the operation differentiable~\cite{balle2016end}.

The generative and the hyper-prior encoder, $g_a$ and $h_a$, are built with the patch merge block and the Visual State Space (VSS) block illustrated in Fig.~\ref{fig:vss}. The patch merge block contains the \textit{Depth-to-Space} operation~\cite{swinT} for down-sampling, a normalization layer, and a linear layer to project the input to a certain depth $C_i$. In $g_a$, the depth $C_i$ of the latent representation increases as the network gets deeper, allowing for a more abstract representation of the image. The size of the latent representation decreases accordingly. In each stage, we down-sample the input feature by $2$. Compared to the convolutional layer used in MambaVC~\cite{qin2024mambavc} for a similar purpose, the chosen patch merge block is easier to implement and is also computationally friendly. 

VSS block, which was originally proposed in~\cite{vmamba},  consists of a single network branch with two residual modules, mimicking the architecture of the vanilla Transformer block~\cite{attention}.  Specifically, each stage in our SSMIC consists of a sequence of VSS blocks and the number of blocks in  stage $i$ is denoted as $d_i$ (see in Fig.~\ref{fig:network}). Given an input feature maps $\mathbf{f} \in \mathbb{R}^{H\times W \times C}$, we get $\mathbf{f''}$ from a first residual module:
\begin{equation}
    \mathbf{f''} = \mathbf{f}+\mathbf{f'},
\end{equation}
where $f'$ is obtained through multiple layers as follows:
\begin{multline}
 \mathbf{f'} = \text{MLP}_2(\text{LN}_2(\text{SS2D}(
    \\
    \sigma( 
    \text{DWConv}(\text{MLP}_1(\text{LN}_1(\mathbb(f)))))))).
\end{multline}
As illustrated in Fig.~\ref{fig:vss} (see (a) and (b)), the output of the VSS block is given by:
\begin{equation}
    \mathbf{f}_{out} =  \text{MLP}_3(\text{LN}_3(\mathbf{f''})) + \mathbf{f''},
\end{equation}
where LN represents the normalization layer; SS2D is the 2D selective scan module; $\sigma$ is the SiLU activation \cite{silo}; DWConv is the depthwise convolution; and MLP is the learnable linear projection. Unlike VSS block in~\cite{vmamba}, we use RMSNorm~\cite{rms} instead of LN LayerNorm since we found empirically that the RMSnorm significantly improves the convergence speed during training.

We adhere to the selective scan approach proposed in~\cite{vmamba} that adapts input-dependent selection mechanism~\cite{mamba} to vision data without reducing its advantages. SS2D involves three steps as shown in Fig.~\ref{fig:vss} (c).
\textit{Cross-scan}: unfolds input features into sequences along four distinct traversal paths; \textit{Selective scan}: processes each path with a distinct S6 in parallel; \textit{Cross-merge}: subsequently reshapes and merges the resultant sequences to form the output feature, enabling effectively integrating information from other pixels in different directions relative to each current pixel. This facilitates the establishment of global receptive fields in the 2D space.

\section{Evaluation and Analysis}
To evaluate the compression efficiency of SSMIC, we propose a performance study across three aspects: rate-distortion (RD), BD-rate, computational complexity and latency.

\begin{table*}[t]
    \centering
    \begin{tabular}{lccccccc}
        \toprule
        Resolution & SSMIC (Ours) &    SwinT \cite{swinT}  & LIC\_TCM \cite{TCM} & ELIC \cite{he2022elic} & LightweightLIC \cite{he2024towards} & MambaVC \cite{qin2024mambavc}& MLIC+ \cite{mlic}\\
        \midrule
        $768\times512$ & \textbf{180.053G} &  208.789G &  215.316G & 231.930G & 239.213G & 326.112G & 452.622G\\
        $1024\times768$ & \textbf{360.106G} &  417.570G & 430.632G & 463.861G & 478.425G & 652.224G & 905.243G\\
        $1280\times1280$ &\textbf{750.222G} & 869.956G  & 897.150G & 966.376G & 996.719G & OM& 1.8859T\\
        \midrule
        \#parameters (M) &  35.79  & 32.34  & 45.18 & 33.79 & 38.39 &53.30 & 83.50\\
        \bottomrule
    \end{tabular}
    \caption{Multiply-Add Cumulation (MACs) for different image resolutions. The last line gives the number of parameters for each model. OM for Out of Memory.}
    \label{tab:macs}
\end{table*}

\begin{table*}[h]
    \centering
    \begin{tabular}{lccccccc}
        \toprule
        Resolution & SSMIC (Ours) &     SwinT \cite{swinT}  & LIC\_TCM \cite{TCM} & ELIC \cite{he2022elic} & LightweightLIC \cite{he2024towards} & MambaVC \cite{qin2024mambavc} & MLIC+ \cite{mlic}\\
        \midrule
        $768\times512$ & 439.618G     & \textbf{419.215G}  & 441.378G & 464.640G& 480.031G & 815.119G & 905.845G \\
        $1024\times768$ & 879.247G & \textbf{838.430G}  & 882.761G & 929.279G & 960.062G & 1.6302T & 1.8117T\\
        $1280\times1280$ &1.8317T &  \textbf{1.7467T} & 1.8391T & 1.9360T &  2.0001T & OM & 3.7744T\\
        \bottomrule
    \end{tabular}
    \caption{Floating Point Operations (FLOPs) for different image resolutions. OM for Out of Memory.}
    \label{tab:flops}
\end{table*}

\subsection{Configuration Setup}
The SSMIC model was trained on the CLIC20 training set~\cite{clic2020}. The loss function used is $L = D + \lambda R$, where $R$ represents the bitrate and $D$ denotes the distortion. The Mean Squared Error
(MSE) in the RGB color space is used as the distortion metric. The Lagrangian multiplier $\lambda$ governs the Rate-Distortion (RD) trade-off. We trained SSMIC with $\lambda \in \{100, 50, 30, 10\}$. Each training batch contains 8 random crops of size $256\times256$. We performed 1 million (1M) iterations using the ADAM optimizer, with the learning rate set to $10^{-4}$. Our SSMIC architecture was implemented in Pytorch using CompressAI library~\cite{compressai}. 

We evaluate the performance of the model on three datasets: Kodak~\cite{kodak}, JPEG-AI~\cite{jpegai}, and the test set of CLIC20~\cite{clic2020} including both mobile and professional categories. During inference, all images were padded with zeros if their size is not a multiple of 256. We compare the proposed SSMIC against the following models: SwinT~\cite{swinT}, MambaVC~\cite{qin2024mambavc}, LightweightLIC~\cite{he2024towards}, LIC\_TCM~\cite{TCM}, SwinNPE~\cite{SwinNPE}, MLIC+~\cite{mlic} and ELIC~\cite{he2022elic}. These models are selected because they are either competitive in terms of compression performance or competitive in terms of the computational efficiency. We also show the compression performance of BPG444~\cite{bpg} as a baseline. For LightweightLIC~\cite{he2024towards}, LIC\_TCM~\cite{TCM}, SwinNPE~\cite{SwinNPE}, and ELIC~\cite{he2022elic}, we evaluated their compression performance and complexity efficiency using their provided pre-trained models under the same configuration as SSMIC. For SwinT~\cite{swinT}, MambaVC~\cite{qin2024mambavc}, and MLIC+~\cite{mlic}, as pre-trained models were not available, we assessed their computational complexity using their untrained models and referenced the RD-curves from their respective papers to evaluate compression performance. The experiments were carried out on an A100 80 Go GPU and an Intel Xeon Gold 6330 3.10 GHz CPU.

\begin{figure}[h]
\vspace{-5mm}
\centering
  \includegraphics[width=0.44\textwidth]{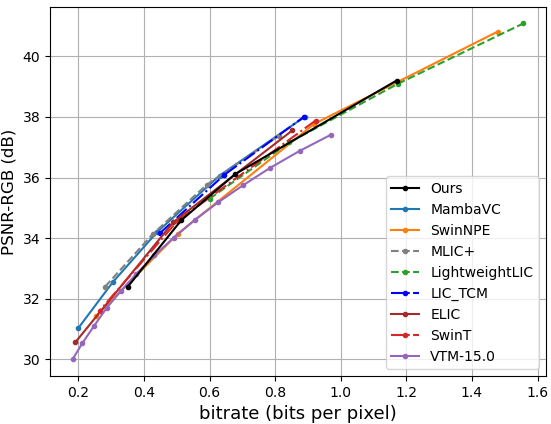}
  \caption{Performance evaluation on the Kodak dataset \cite{kodak}.}
  \label{fig:RD_curve}
\end{figure}

\subsection{Performance Comparison}
We summarize in Table~\ref{tab:compression_comparison}, the BD-rate of SSMIC and the competitive state-of-the-art models across three datasets. The BD-rate was calculated covering approximately 0.4 to 1.2 bits per pixel (bpp), using VTM-15.0~\cite{VTM} as a reference. 

On average, SSMIC achieves a -21.75\% BD-rate performance compared to VTM-15.0 and $4.17$ p.p., relative
increase from LIC\_TCM \cite{TCM} which provides the best performance compared to the other selected methods.  However, the latter is much more demanding in terms of computational 
complexity and number of parameters. This is shown in Fig.~\ref{fig:bubble}, which provides the BD-rate with VTM-15.0 as an anchor versus the GMACs~\footnote[1]
{MACs and FLOPs are calculated using calflops Library 
\url{https://github.com/MrYxJ/calculate-flops.pytorch} } 
of various approaches on the Kodak dataset. These observations confirm that our SSMIC offers a good trade-off between BD-rate performance and computational complexity.

We also evaluate the compression performance of different models across various bitrate ranges on the Kodak dataset in Fig.~\ref{fig:RD_curve}. The results show that the proposed model consistently achieves performance comparable to state-of-the-art methods while maintaining competitive computational efficiency. Indeed, we show in Tables~\ref{tab:macs} and~\ref{tab:flops} the computational complexity in terms of MACs\footnotemark[1] and FLOPs\footnotemark[1], respectively.  The results are shown for three different resolutions, selected for their common usage and to avoid out-of-memory issues on the utilized GPU. It is clear that our SSMIC significantly reduces the computational complexity in terms of MACs and achieves competitive results in terms of FLOPs compared to SwinT~\cite{swinT}. 

Table \ref{tab:latency} presents the average latency over 2,000 images at a resolution of $256 \times 256$ on the utilized GPU. Our model shows competitive decoding times compared to lightweightLIC~\cite{he2024towards}, while maintaining comparable encoding times.

\begin{table}[h]
    \centering
    \begin{tabular}{lcc}
        \toprule
        Method & \multicolumn{2}{c}{Latency  GPU (ms)} \\
        \cmidrule(lr){2-3} & Encoding & Decoding \\
        \midrule
        ELIC \cite{he2022elic}& 426.94& 517.15\\
        MambaVC \cite{qin2024mambavc}& \textbf{14.01} & 73.36 \\
        LIC\_TCM  \cite{TCM}& 15.23 & 52.46 \\
        SwinT \cite{swinT}& 14.25 & 20.86  \\
        LightweightLIC \cite{he2024towards}& 15.62 & \textbf{15.56} \\
        SSMIC (Ours)  & 20.24 & 19.93 \\ 
        \bottomrule
    \end{tabular}
    \caption{Average latency,  measured on an 
A100 80 Go GPU and an Intel Xeon Gold 6330
3.10 GHz CPU,  using over 2000 images at $256 \times 256$ resolution.}
    \label{tab:latency}
\end{table}

\section{Conclusion}
In this paper, we  proposed State Space Model-based Image Compression (SSMIC) approach, achieving competitive RD performance while significantly reducing the computational complexity and latency, which is potentially helpful to conduct, with further optimizations, high-quality real-time visual data compression. SSMIC leverages the advantages of State Space Models (SSMs) inherited from the Mamba model, enhancing contextual
reasoning while managing computational and memory requirements.  On average across benchmark datasets, SSMIC achieves a substantial reduction in BD-rate compared to VTM-15.0, underscoring its effectiveness in practical applications.

\bibliographystyle{Efficient_Image_Compression_Using_Advanced_State_Space_Models}
\bibliography{Efficient_Image_Compression_Using_Advanced_State_Space_Models}

\begin{thebibliography}{10}

\bibitem{balle2018variational}
Johannes Ball{\'e}, David Minnen, Saurabh Singh, Sung~Jin Hwang, and Nick Johnston,
\newblock ``Variational image compression with a scale hyperprior,''
\newblock in {\em International Conference on Learning Representations}, 2018.

\bibitem{Learning_Convo}
Mu~Li, Wangmeng Zuo, Shuhang Gu, Debin Zhao, and David Zhang,
\newblock ``Learning convolutional networks for content-weighted image compression,''
\newblock in {\em Proceedings of the IEEE conference on computer vision and pattern recognition}, 2018, pp. 3214--3223.

\bibitem{CondiProModel}
Fabian Mentzer, Eirikur Agustsson, Michael Tschannen, Radu Timofte, and Luc Van~Gool,
\newblock ``Conditional probability models for deep image compression,''
\newblock in {\em Proceedings of the IEEE Conference on Computer Vision and Pattern Recognition}, 2018, pp. 4394--4402.

\bibitem{NEURIPS2018_53edebc5}
David Minnen, Johannes Ball{\'e}, and George~D Toderici,
\newblock ``Joint autoregressive and hierarchical priors for learned image compression,''
\newblock {\em Advances in neural information processing systems}, vol. 31, 2018.

\bibitem{lee2018contextadaptive}
Jooyoung Lee, Seunghyun Cho, and Seung-Kwon Beack,
\newblock ``Context-adaptive entropy model for end-to-end optimized image compression,''
\newblock {\em arXiv preprint arXiv:1809.10452}, 2018.

\bibitem{9190935}
David Minnen and Saurabh Singh,
\newblock ``Channel-wise autoregressive entropy models for learned image compression,''
\newblock in {\em 2020 IEEE International Conference on Image Processing (ICIP)}. IEEE, 2020, pp. 3339--3343.

\bibitem{Entroformer}
Yichen Qian, Ming Lin, Xiuyu Sun, Zhiyu Tan, and Rong Jin,
\newblock ``Entroformer: A transformer-based entropy model for learned image compression,''
\newblock {\em arXiv preprint arXiv:2202.05492}, 2022.

\bibitem{swinT}
Yinhao Zhu, Yang Yang, and Taco Cohen,
\newblock ``Transformer-based transform coding,''
\newblock in {\em International Conference on Learning Representations}, 2021.

\bibitem{stf}
Renjie Zou, Chunfeng Song, and Zhaoxiang Zhang,
\newblock ``The devil is in the details: Window-based attention for image compression,''
\newblock in {\em Proceedings of the IEEE/CVF conference on computer vision and pattern recognition}, 2022, pp. 17492--17501.

\bibitem{beyer2022knowledge}
Lucas Beyer, Xiaohua Zhai, Am{\'e}lie Royer, Larisa Markeeva, Rohan Anil, and Alexander Kolesnikov,
\newblock ``Knowledge distillation: A good teacher is patient and consistent,''
\newblock in {\em Proceedings of the IEEE/CVF conference on computer vision and pattern recognition}, 2022.

\bibitem{guo2022evc}
Wang Guo-Hua, Jiahao Li, Bin Li, and Yan Lu,
\newblock ``Evc: Towards real-time neural image compression with mask decay,''
\newblock in {\em The Eleventh International Conference on Learning Representations}, 2022.

\bibitem{wu2023pslt}
Gaojie Wu, Wei-Shi Zheng, Yutong Lu, and Qi~Tian,
\newblock ``Pslt: a light-weight vision transformer with ladder self-attention and progressive shift,''
\newblock {\em IEEE Transactions on Pattern Analysis and Machine Intelligence}, 2023.

\bibitem{mehta2021mobilevit}
Sachin Mehta and Mohammad Rastegari,
\newblock ``Mobilevit: light-weight, general-purpose, and mobile-friendly vision transformer,''
\newblock {\em arXiv preprint arXiv:2110.02178}, 2021.

\bibitem{he2024towards}
Ziyang He, Minfeng Huang, Lei Luo, Xu~Yang, and Ce~Zhu,
\newblock ``Towards real-time practical image compression with lightweight attention,''
\newblock {\em Expert Systems with Applications}, vol. 252, pp. 124142, 2024.

\bibitem{yang2023computationally}
Yibo Yang and Stephan Mandt,
\newblock ``Computationally-efficient neural image compression with shallow decoders,''
\newblock in {\em Proceedings of the IEEE/CVF International Conference on Computer Vision}, 2023, pp. 530--540.

\bibitem{he2021checkerboard}
Dailan He, Yaoyan Zheng, Baocheng Sun, Yan Wang, and Hongwei Qin,
\newblock ``Checkerboard context model for efficient learned image compression,''
\newblock in {\em Proceedings of the IEEE/CVF Conference on Computer Vision and Pattern Recognition}, 2021, pp. 14771--14780.

\bibitem{ali2024towards}
Muhammad~Salman Ali, Yeongwoong Kim, Maryam Qamar, Sung-Chang Lim, Donghyun Kim, Chaoning Zhang, Sung-Ho Bae, and Hui~Yong Kim,
\newblock ``Towards efficient image compression without autoregressive models,''
\newblock {\em Advances in Neural Information Processing Systems}, vol. 36, 2024.

\bibitem{VTM}
Benjamin Bross, Ye-Kui Wang, Yan Ye, Shan Liu, Jianle Chen, Gary Sullivan, and Jens-Rainer Ohm,
\newblock ``Overview of the versatile video coding (vvc) standard and its applications,''
\newblock {\em IEEE Transactions on Circuits and Systems for Video Technology}, vol. 31, pp. 3736--3764, 10 2021.

\bibitem{kodak}
Rich Franzen,
\newblock ``Kodak lossless true color image suite,''
\newblock \url{ http://r0k. us/graphics/kodak}, 1999.

\bibitem{ssm}
Albert Gu, Isys Johnson, Karan Goel, Khaled Saab, Tri Dao, Atri Rudra, and Christopher R{\'e},
\newblock ``Combining recurrent, convolutional, and continuous-time models with linear state space layers,''
\newblock {\em Advances in neural information processing systems}, vol. 34, pp. 572--585, 2021.

\bibitem{s4}
Albert Gu, Karan Goel, and Christopher R{\'e},
\newblock ``Efficiently modeling long sequences with structured state spaces,''
\newblock {\em arXiv preprint arXiv:2111.00396}, 2021.

\bibitem{liu2021swin}
Ze~Liu, Yutong Lin, Yue Cao, Han Hu, Yixuan Wei, Zheng Zhang, Stephen Lin, and Baining Guo,
\newblock ``Swin transformer: Hierarchical vision transformer using shifted windows,''
\newblock in {\em Proceedings of the IEEE/CVF international conference on computer vision}, 2021, pp. 10012--10022.

\bibitem{TCM}
Jinming Liu, Heming Sun, and Jiro Katto,
\newblock ``Learned image compression with mixed transformer-cnn architectures,''
\newblock in {\em Proceedings of the IEEE/CVF Conference on Computer Vision and Pattern Recognition}, 2023, pp. 14388--14397.

\bibitem{chen2022mobile}
Yinpeng Chen, Xiyang Dai, Dongdong Chen, Mengchen Liu, Xiaoyi Dong, Lu~Yuan, and Zicheng Liu,
\newblock ``Mobile-former: Bridging mobilenet and transformer,''
\newblock in {\em Proceedings of the IEEE/CVF conference on computer vision and pattern recognition}, 2022, pp. 5270--5279.

\bibitem{mobilenet}
Andrew~G. Howard, Menglong Zhu, Bo~Chen, Dmitry Kalenichenko, Weijun Wang, Tobias Weyand, Marco Andreetto, and Hartwig Adam,
\newblock ``Mobilenets: Efficient convolutional neural networks for mobile vision applications,'' 2017.

\bibitem{pan2022edgevits}
Junting Pan, Adrian Bulat, Fuwen Tan, Xiatian Zhu, Lukasz Dudziak, Hongsheng Li, Georgios Tzimiropoulos, and Brais Martinez,
\newblock ``Edgevits: Competing light-weight cnns on mobile devices with vision transformers,''
\newblock in {\em European Conference on Computer Vision}. Springer, 2022.

\bibitem{maaz2022edgenext}
Muhammad Maaz, Abdelrahman Shaker, Hisham Cholakkal, Salman Khan, Syed~Waqas Zamir, Rao~Muhammad Anwer, and Fahad Shahbaz~Khan,
\newblock ``Edgenext: efficiently amalgamated cnn-transformer architecture for mobile vision applications,''
\newblock in {\em European Conference on Computer Vision}. Springer, 2022, pp. 3--20.

\bibitem{yang2022lite}
Chenglin Yang, Yilin Wang, Jianming Zhang, He~Zhang, Zijun Wei, Zhe Lin, and Alan Yuille,
\newblock ``Lite vision transformer with enhanced self-attention,''
\newblock in {\em Proceedings of the IEEE/CVF Conference on Computer Vision and Pattern Recognition}, 2022, pp. 11998--12008.

\bibitem{zhang2022efficient}
Xindong Zhang, Hui Zeng, Shi Guo, and Lei Zhang,
\newblock ``Efficient long-range attention network for image super-resolution,''
\newblock in {\em European conference on computer vision}. Springer, 2022, pp. 649--667.

\bibitem{SwinNPE}
Bouzid Arezki, Fangchen Feng, and Anissa Mokraoui,
\newblock ``Convolutional transformer-based image compression,''
\newblock in {\em 2023 Signal Processing: Algorithms, Architectures, Arrangements, and Applications (SPA)}. IEEE, 2023, pp. 154--159.

\bibitem{qin2024mambavc}
Shiyu Qin, Jinpeng Wang, Yimin Zhou, Bin Chen, Tianci Luo, Baoyi An, Tao Dai, Shutao Xia, and Yaowei Wang,
\newblock ``Mambavc: Learned visual compression with selective state spaces,''
\newblock {\em arXiv preprint arXiv:2405.15413}, 2024.

\bibitem{vmamba}
Yue Liu, Yunjie Tian, Yuzhong Zhao, Hongtian Yu, Lingxi Xie, Yaowei Wang, Qixiang Ye, and Yunfan Liu,
\newblock ``Vmamba: Visual state space model,''
\newblock {\em arXiv preprint arXiv:2401.10166}, 2024.

\bibitem{mamba}
Albert Gu and Tri Dao,
\newblock ``Mamba: Linear-time sequence modeling with selective state spaces,''
\newblock {\em arXiv preprint arXiv:2312.00752}, 2023.

\bibitem{liu2024vision}
Xiao Liu, Chenxu Zhang, and Lei Zhang,
\newblock ``Vision mamba: A comprehensive survey and taxonomy,''
\newblock {\em arXiv preprint arXiv:2405.04404}, 2024.

\bibitem{clic2020}
Wenzhe~Shi George~Toderici, Radu Timofte, Lucas Theis, Johannes Balle, Eirikur Agustsson, Nick Johnston, and Fabian Mentzer,
\newblock ``Workshop and challenge on learned image compression (clic2020),'' 2020.

\bibitem{jpegai}
JPEG-AI,
\newblock ``Jpeg-ai test images,''
\newblock \url{https://jpegai.github.io/test\_images/}, 2020.

\bibitem{bpg}
Fabrice Bellard,
\newblock ``Bpg image format,''
\newblock \url{ http://bellard.org/bpg/}, 2018.

\bibitem{he2022elic}
Dailan He, Ziming Yang, Weikun Peng, Rui Ma, Hongwei Qin, and Yan Wang,
\newblock ``Elic: Efficient learned image compression with unevenly grouped space-channel contextual adaptive coding,''
\newblock in {\em Proceedings of the IEEE/CVF Conference on Computer Vision and Pattern Recognition}, 2022, pp. 5718--5727.

\bibitem{mlic}
Wei Jiang, Jiayu Yang, Yongqi Zhai, Peirong Ning, Feng Gao, and Ronggang Wang,
\newblock ``Mlic: Multi-reference entropy model for learned image compression,''
\newblock in {\em Proceedings of the 31st ACM International Conference on Multimedia}. Oct. 2023, MM ’23, ACM.

\bibitem{balle2016end}
Johannes Ball{\'e}, Valero Laparra, and Eero~P Simoncelli,
\newblock ``End-to-end optimization of nonlinear transform codes for perceptual quality,''
\newblock in {\em 2016 Picture Coding Symposium (PCS)}. IEEE, 2016, pp. 1--5.

\bibitem{attention}
Ashish Vaswani, Noam Shazeer, Niki Parmar, Jakob Uszkoreit, Llion Jones, Aidan~N Gomez, {\L}ukasz Kaiser, and Illia Polosukhin,
\newblock ``Attention is all you need,''
\newblock {\em Advances in neural information processing systems}, vol. 30, 2017.

\bibitem{silo}
Prajit Ramachandran, Barret Zoph, and Quoc~V Le,
\newblock ``Searching for activation functions,''
\newblock {\em arXiv preprint arXiv:1710.05941}, 2017.

\bibitem{rms}
Biao Zhang and Rico Sennrich,
\newblock ``Root mean square layer normalization,''
\newblock {\em Advances in Neural Information Processing Systems}, vol. 32, 2019.

\bibitem{compressai}
Jean Bégaint, Fabien Racapé, Simon Feltman, and Akshay Pushparaja,
\newblock ``Compressai: a pytorch library and evaluation platform for end-to-end compression research,'' 2020.

\end{thebibliography}

\end{document}